\documentclass{pazhastl}
\usepackage{graphicx}
\usepackage{graphicx,natbib}
\bibpunct{(}{)}{;}{a}{}{,}

\def\smfigurewocap#1#2#3{
  \begin{minipage}{1.0\columnwidth}
    \begin{minipage}{0.049\columnwidth}
      \rotatebox{90}{\phantom{0000}#3}
    \end{minipage}
    \begin{minipage}{0.9\columnwidth}
      \includegraphics[bb=58 188 556 678,width=0.97\columnwidth]{#1}
      \centerline{#2}
    \end{minipage}
    
    \vskip 3pt
    ~
  \end{minipage}
}

\begin{document}

\journalinfo{2008}{0}{0}{1}[0] 

\title{Optical identification of hard X-ray source IGR\,J18257$-$0707}

\author{
  R.~A.~Burenin\email{rodion@hea.iki.rssi.ru}\address{1},
  I.~F.~Bikmaev\address{2,3},
  M.~G.~Revnivtsev\address{1,4}, 
  J.~A.~Tomsick\address{5},\\
  S.~Yu.~Sazonov\address{1,4},
  M.~N.~Pavlinskiy\address{1},
  R.~A.~Sunyaev\address{1,4}
  \addresstext{1}{Space Research Institute (IKI), Moscow, Russia}
  \addresstext{2}{Kazan State University, Kazan, Russia}
  \addresstext{3}{Tatarstan Academy of Sciences, Kazan, Russia}
  \addresstext{4}{Max-Planck-Instutut fuer Astrophysik, Garching, Germany}
  \addresstext{5}{Space Sciences Laboratory, University of California,
    Berkeley, CA, USA}
}

\shortauthor{Burenin et al.}

\shorttitle{Optical identification of IGR\,J18257$-$0707}
\submitted{June 23, 2008}

\begin{abstract}
  We present the results of the optical identification of hard X-ray source
  IGR\,J18257$-$0707 trough the spectroscopic observations of its optical
  counterpart with RTT150 telescope. Accurate position of the X-ray source,
  determined using Chandra observations, allowed us to associate this source
  with the faint optical object ($m_R\approx 20.4$), which shows broad
  $H_\alpha$ emission line in its optical spectrum.  Therefore we conclude
  that the source IGR\,J18257$-$0707 is a type 1 Seyfert galaxy at redshift
  $z=0.037$.
  
\keywords{X-ray sources  --- gamma-sources --- active galactic nuclei
    --- optical observations}

\end{abstract}

The hard X-ray source IGR\,J18257$-$0707 was discovered in INTEGRAL all sky
survey \citep{krivonos07}. This source is also known in literature as
IGR\,J18259$-$0706 \citep[e.g.,][]{stephen06}. Our group systematically
observe unidentified INTEGRAL sources with the aim to determine their nature
\citep{bikmaev2006a,bikmaev2006b,bikmaev2008,burenin08,sazonov08,knyazev08}.
In the case of IGR\,J18257$-$0707 the association of the X-ray source with
its optical counterpart was complicated, therefore we present the results of
the optical identification of this source in this separate Letter.

\begin{figure}
  \centering
  \includegraphics[width=0.9\columnwidth]{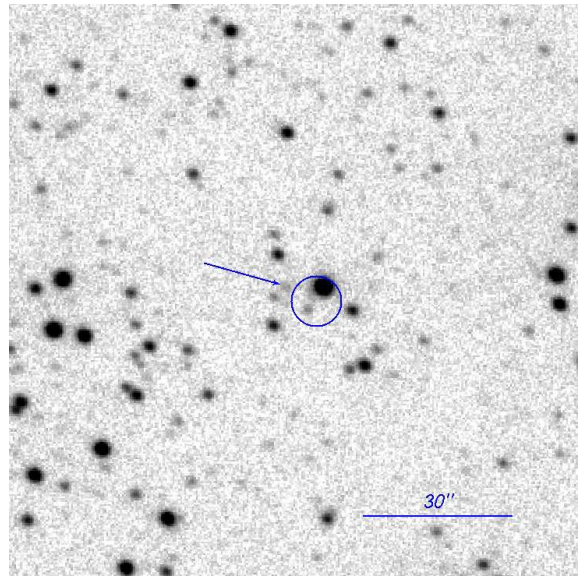}
  \caption{The image of the field near IGR\,J18257$-$0707 in \emph{R} band.
    The position of the source determined with SWIFT/XRT is shown with the
    circle. The size of the circle correspond to the SWIFT/XRT localization
    uncertainty (6.5\arcsec). Arrow denote the object which was identified
    as an optical counterpart of the X-ray source using Chandra
    observations.}
  \label{fchart}
\end{figure}

\begin{figure*}
  \centering 
  \smfigurewocap{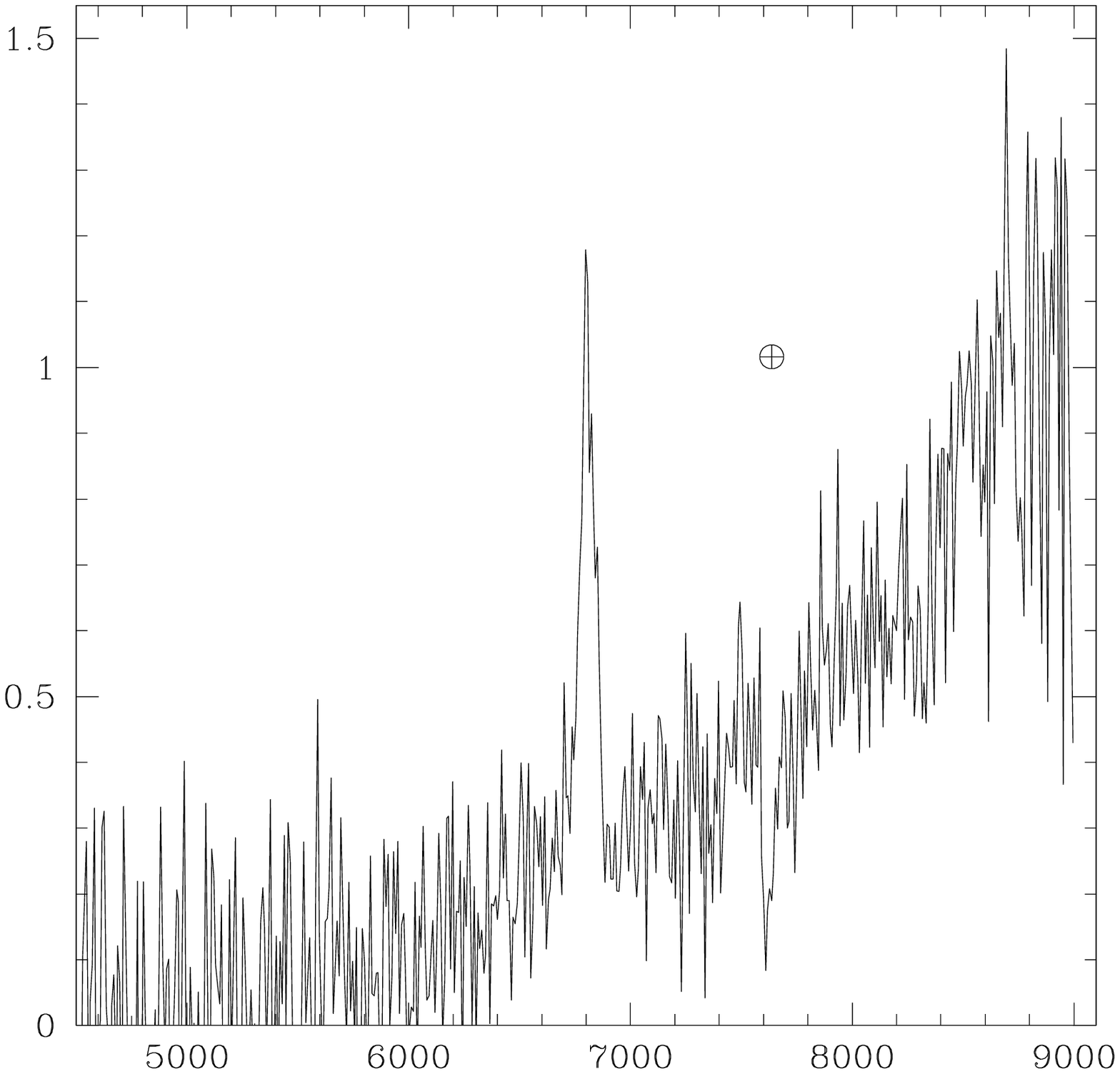}{$\lambda$,
    \AA}{$F_\lambda, \times 10^{-16}$~erg~s$^{-1}$~cm$^{-2}$~\AA$^{-1}$}
  \smfigurewocap{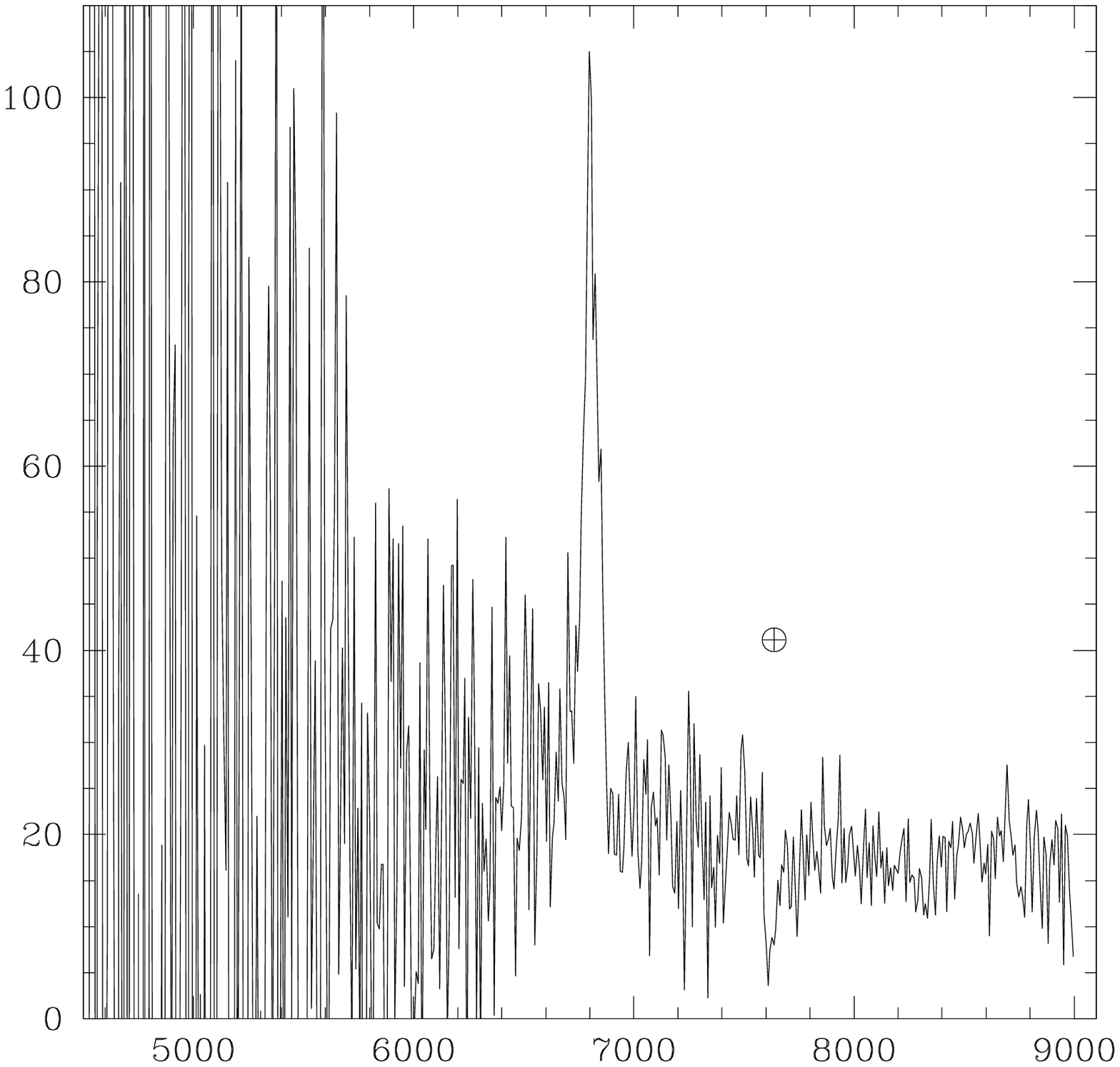}{$\lambda$,
    \AA}{$F_\lambda, \times 10^{-16}$~erg~s$^{-1}$~cm$^{-2}$~\AA$^{-1}$}
  \caption{Left: the spectrum of the optical object, associated with hard
    X-ray source IGR\,J18257$-$0707. Right: the same spectrum, corrected for
    the Galactic extinction $E(B-V)=2.01$.}
  \label{fig:sp}
\end{figure*}

In soft X-rays the source was found also in ROSAT All-Sky Survey (RASS) and
is included in ROSAT Bright Source Catalog as 1RXS\,J182557.5$-$071021
\citep{voges99}. The error circle for this source is of $\approx20\arcsec$
radius --- much more accurate than the localization error of INTEGRAL hard
X-ray source, which is $\approx5\arcmin$. However, this improved accuracy is
still insufficient to associate this source with some object in optical band
(Fig.~\ref{fchart}).

Position of the hard X-ray source was refined using the X-ray telescope
aboard SWIFT satellite, observations were made on Feb.\ 12, 2007. These data
give the following coordinates of the X-ray source: $\alpha,\delta
=$18:25:57.12, $-$07:10:24.4 (J2000). Astrometric uncertainty of this
position is about 6.5\arcsec\ and is mainly systematic \citep{moretti06}.
Corresponding error circle is shown in Fig.~\ref{fchart}.

The object is located at low Galactic latitude, $b=2.35^\circ$, therefore
due to high stellar density in this area there are more than one optical
objects in SWIFT/XRT error circle. Spectra of some of these objects were
obtained with Russian-Turkish telescope (RTT150). In particular, using
\emph{TFOSC} spectrometer we obtained the spectrum of the bright
($m_R\approx14.6$) star, located within the circle (Fig.~\ref{fchart}). This
object appeared to be an ordinary \emph{K}-type star and there is a fair
probability for such a star to be located in the SWIFT/XRT error circle by
chance.

In order to obtained more accurate position of the X-ray source we used
Chandra observations of this field made on Feb.14, 2008. With these data the
position of the source was determined with the accuracy 0.64\arcsec\ (90\%
confidence): $\alpha,\delta =$18:25:57.58, $-$07:10:22.8 (J2000). The
detailed results of Chandra observations of the source will be presented in
a separate paper \citep{tomsick08}.

Refined position of the source IGR\,J18257$-$0707 allowed to associate it
with the optical object, shown with arrow in Fig.~\ref{fchart}. This object
is located in about 7\arcsec\ from the center of SWIFT/XRT error circle, the
position of the optical object also coincides with the position of infrared
source 2MASS\,J18255759$-$0710229 from 2MASS survey \citep{cutri03}.

The magnitude of the optical counterpart of IGR\,J18257$-$0707 is
$m_R\approx 20.4$ and usually objects with similar magnitudes are considered
to be to faint for spectroscopic observations with 1.5-m telescope. However,
even for those faint objects, 1.5-m telescope is able to detect bright
emission lines if they are present in the spectrum of the source. With the
aim to search for these bright emission lines, with RTT150 telescope we
obtained a number of spectra of the source with 6300~s total exposure time.

In these observations we used the 100~$\mu$m slit which corresponds to
1.78\arcsec\ in the sky and grism \#15 which provide the best optical
transmission and the most wide spectral coverage (3500--9000~\AA). Spectral
resolution with this setup is $\approx15$~\AA\ (FWHM). The data were reduced
using standard \emph{IRAF}\footnote{http://iraf.noao.edu} software. The
aperture for the extraction of the object spectrum was defined using the
spectrum of the nearby bright star which was also found in the slit.

The spectrum obtained in these observations is shown in Fig.~\ref{fig:sp}.
One can see the bright and broad emission line in the spectrum. The
equivalent width of this line is $-276$~\AA, full width at half maximum is
$3600$~km~s$^{-1}$. The lines with that large width are known only as
hydrogen lines in quasars and AGNs spectra. Also, that can not be a line,
where the other nearby bright lines are detected. For example, near the
$H\beta$ line one should detect the bright \mbox{[OIII], 4959,5007}
forbidden narrow emission lines, near the $L\alpha$ line bright NV,
SiIV+OIV, CIV lines should be detected.

We conclude that this line should be identified with $H\alpha$ line at
redshift $z=0.037$.  In that case the object is identified as nearby type 1
Seyfert galaxy, which constitute the main part of extragalactic objects
among the optical identifications of new INTEGRAL hard X-ray sources
\citep[see, e.g.][]{bikmaev2006a,burenin08}. The other emission lines which
are usually observed in Seyfert 1 galaxies are not detected here for the
following reasons: bright $H_\beta$ and [OIII] lines are in blue part of the
spectrum which is heavily absorbed in Galaxy.  Other lines near $H_\alpha$,
like \mbox{[SII]} are much weaker than $H_\alpha$ and are not detected due
to low signal-to-noise ratio in our data.

\begin{figure}
  \centering
  \includegraphics[width=0.85\columnwidth]{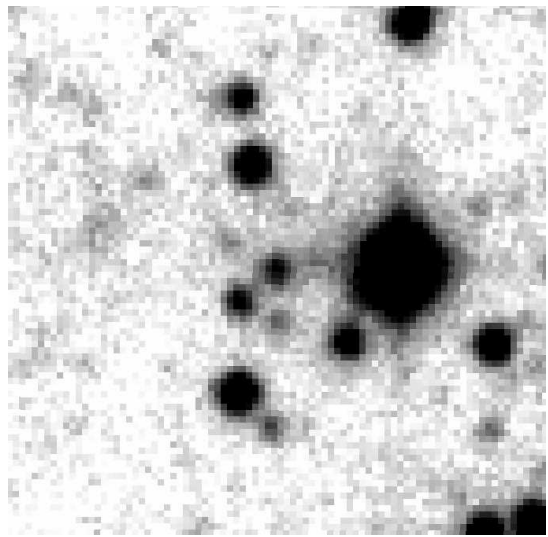}
  \caption{The image of AGN identified with hard X-ray source
    IGR\,J18257$-$0707 in filter \emph{I}.}
  \label{fig:i}
\end{figure}

Spectrum corrected for interstellar reddening \mbox{$E(B-V)=2.01$}
\citep{schlegel98} is also shown in Fig.~\ref {fig:sp}. The overall shape of
dereddened spectrum is consistent with the presence of nonthermal continuum
which is typically observed in optical spectra of Seyfert galaxies
\citep[e.g.,][]{elvis94}. The interstellar reddening $E(B-V)=2.01$
approximately corresponds to $A(R)\approx5.1$, therefore the brightness of
the object corrected for Galactic absorption should be $\approx15.3$. If the
AGN host galaxy were of comparable magnitude, it would be detected in the
optical images as an extended source. However, in our case we could not
reliably determine the object extent, because the observations are hampered
by PSF wings from the bright star in vicinity of the object.  This is
illustrated in Fig.~\ref{fig:i}.

The line of sight absorption column, measured in the X-ray spectrum of
IGR\,J18257$-$0707 with SWIFT is approximately $n_H L =
(1.4\pm0.4)\times10^{22}$~cm$^{-2}$. From the other hand, the interstellar
reddening $E(B-V)=2.01$ with typical ratio $n_H L/E(B-V)\approx
4.8\times10^{21}$~cm$^2$ corresponds to the line of sight absorption column
about $n_H L \approx1\times10^{22}$ cm$^{-2}$.  Approximately the same value
$n_H L\approx0.7\times10^{22}$ cm$^{-2}$ can be obtained from the HI map of
the Galaxy \citep{dickey90}. Therefore, our data do not contradict to the
suggestion that the absorption observed in X-rays originate completely in
the disk of our Galaxy.

Thus, in result of our study of the hard X-ray source IGR\,J18257$-$0707 we
conclude that this source is an active galactic nucleus --- type 1 Seyfert
galaxy at redshift $z=0.037$.

\acknowledgements

Our work is supported by grants RFFI 07-02-01004, RFFI 08-02-00974,
NSH-4224.2008.2 and NSH-5579.2008.2, and also by programs of Russian Academy
of Sciences P-04 and OFN-17.

\end{document}